# A Theory for the Cosmological Constant and its Explanation of the Gravitational Constant

H.M.Mok

Radiation Health Unit, 3/F., Saiwanho Health Centre,

Hong Kong SAR Govt, 28 Tai Hong St.,

Saiwanho, Hong Kong, China.

e-mail: jhmmok@netvigator.com

**Abstract** This paper continues the development of a discrete space-time concept that is recently used in the explanation of the cosmological constant. Instead of order estimation, a more theoretical treatment of the theory is introduced. Based on the discrete space-time hypothesis, we find that the electroweak theory and General Relativity can be combined into a single theory. An analytical form of the cosmological constant can be found and the weakness problem of the gravitational constant can also be explained.

PACS numbers: 98.80.Cq, 04.60.-m

In a recent theory [1], we proposed that the mystery of the cosmological constant problem (some recent review on the problem can be found in [2,3,4]) can be resolved by the assumption of discrete space-time. Although G.'t Hooft had proposed that a discrete nature of space-time can be derived by combining certain features of General Relativity (GR) and quantum mechanics [5], we treat it as a hypothesis for simplicity. Based on the following three hypotheses: (1) Space-time is discrete in nature and its fundamental unit is of the order of Planck scale; (2) The space-time forms a kind of phase (or say "condensate") with its constituents; (3) The scalar field in electroweak theory plays both the role as the order parameter of such space-time phase and the wavefunction of its constituents, it is shown that space-time is discrete in the electroweak scale. The cosmological constant estimated by the theory is in excellent agreement with the Type Ia SN observations [6,7]. The most recent analysis of SN1997ff also confirms such results [8]. Besides, the theory has important implication on the problems other than the cosmological constant. Firstly, cosmic inflation is just a consequence of the theory. Secondly, the universe is found to be alternatively dominated by the cosmological constant and the mass density at different cosmic time period. Our calculation also shows that the matter energy density $\rho_m$ is of similar order of magnitude as the vacuum energy density $\rho_v$ in the present universe but it is just a coincidence. This result supports the anthropic principle [9]. The theory of discrete space-time also opens a new direction in quantum gravity. It is because if the space-time is made of more fundamental unit or constituents, the divergence problem of the quantum gravity would be of no interest since $g_{\mu\nu}$ is not a fundamental field and General Relativity is just a collective effect of the lattice like vibration of the space-time constituents. In this paper, we continue the development of the discrete space-time concept and try to give mathematical description on the theory and construct a more detail model on the space-time microscopic structure. The value of the gravitational constant and its weakness are also explained.

If the space-time manifold is made of more fundamental constituents which are subjected to quantum mechanical effects, a new definition of the metric may be needed to describe both the quantum behaviour of the constituents in microscopic scale and their macroscopic effect as a smooth manifold. Let us define a new generalized metric that involves the contribution of a $\phi$ field as

$$g'_{\mu\nu}(x) = g_{\mu\nu}(x)\phi^2(x) \qquad (1)$$

where $\phi(x)$ is the field "related to" the wavefunction of the constituent (later we will show that the $\phi$ field is not exactly equal to the wavefunction of the space-time constituent but the "holes" of it does). $g_{\mu\nu}$ is defined to be the metric which is independent to the $\phi$ field. Although the mathematical form of equation (1) is similar to the conformal transformation on $g_{\mu\nu}$, we treat $\phi$ as a real physical field in our theory rather than just a conformal function. The introduction of the $\phi$ field in the metric here will create some new space-time concepts. Firstly, if the symmetry of the $\phi$ field is spontaneously broken and its vacuum expectation value (VEV) takes a constant value, the distribution of the space-time constituents of the manifold is uniform. Secondly, different symmetry broken stages of $\phi$ forms different "phase" of the space-time in which the density of the constituents may be different, as described in the previous paper [1]. The space-time of the present universe could be just in one of the broken phase of $\phi$ since the density of the space-time constituents is expected to be uniform. Using these concepts, we will later show that GR and electroweak theory can be combined into a single theory such that the vacuum state described in electroweak theory is just the "phase" of the space-time and its transition properties is determined by the $\phi$ field.

Let us assume the conformal transformation of $g_{\mu\nu}$ and $\phi$ as $g_{\mu\nu} \to \Omega^2 g_{\mu\nu}$ and $\phi \to \Omega^{-1}\phi$. $g'_{\mu\nu}$ in equation(1) is then a conformal invariant and the scale of $g_{\mu\nu}$ is not determined before fixing the value of $\phi$. If we put $g'_{\mu\nu}$of equation (1) into the usual GR action, a generalized conformal invariant action can be obtained as

$$S_g = \frac{-1}{2\kappa}\int (R'-2\lambda)\sqrt{-g'}d^4x \qquad (2)$$

where $R'$ is the Ricci scalar corresponding to our new metric $g'_{\mu\nu}$. After simple calculation, we get

$$S_g = \frac{-1}{2\kappa}\int (R(x)\phi^2 + 6\partial_\mu\phi\partial^\mu\phi - 2\lambda\phi^4)\sqrt{-g}d^4x \qquad (3)$$

where $R(x)$ is the Ricci scalar corresponding to the metric $g_{\mu\nu}$. If we varies the action with respect to $g_{\mu\nu}$, a conformal invariant equation can be obtained. Through that equation, when $\phi$ takes a constant value, for example, $\phi_0$, the coupling constant becomes $\kappa / \phi^2$ and the relationship between $R$ & $\lambda$ can be found as $R = 4\lambda\phi_0{}^2$. The cosmological constant is equivalent to the term $\lambda\phi_0^4$. If we change the variable of $\phi$ as $\phi = \psi\sqrt{\kappa/6}$, equation (3) becomes

$$S_g = -\int (\frac{1}{2}\partial_\mu \psi \partial^\mu \psi + \frac{1}{12}R\psi^2 - \frac{1}{36}\lambda\kappa\psi^4)\sqrt{-g}d^4x \qquad (4)$$

Now, if we make use of the above hypothesis (1) that the space-time manifold is made of some fundamental unit, we expect that $R$ would tend to be a constant $R_0$ and be independent with $g_{\mu\nu}$ when the space-time is about that length scale. This situation is similar to a Ricci scalar defined on an elastic membrane with atomic structure. The Ricci scalar describes the curvature of the membrane and is a function of the metric macroscopically but it decouple with the metric in atomic scale due to the domination of curvature by individual atom. On fixing the Ricci scalar, the conformal symmetry of the action will be broken and the scale of the metric can be fixed. However, one may find that the metric $g_{\mu\nu}$ cannot be determined solely by the action in equation (4) afterwards and it should be find by the macroscopic equation. That means we have to build the metric in macroscopic scale in order to describe the field of the microscopic constituent. This situation can be analogous to the description of the phonon propagation on a curved membrane and, in general, the case is not difficult for the fundamental scale of the space-time since the background metric can be assumed to be locally flat and therefore it is of Minkowski type in nature. In obtaining the macroscopic equation, $R$ is still a function of the metric and the action is conformal invariant. If we fix the coupling strength of metric and break the conformal symmetry by choosing $\phi = 1$, the action becomes the one commonly used in GR so that the metric can be found. (We will later show that fixing the coupling strength is directly related to fixing the size of the constituents). The conformal symmetry broken in different scale does not pose us problem here because, in general, the symmetry broken in microscopic scale can be approximately retained in macroscopic scale due to the smoothing out of large amount of microscopic particle. This can be analogous to the continuous translation symmetry of an infinite atomic lattice. Such symmetry is broken in atomic scale but approximately be retained in macroscopic scale. So that, we can further break the macroscopic symmetry by some other means. Therefore, the translation symmetry is doubly broken in macroscopic and microscopic scale. The above argument is an important step that we can combine GR with electroweak theory by the discrete space-time hypothesis because the action in equation (3) converge in microscopic scale to the lagrangian for a scalar field, that can be identified as the one used in electroweak theory, and it becomes the GR action in macroscopic scale. However, the lagrangian differs from the usual electroweak theory by an overall negative sign. This problem can be solved by using the concept of holes in the solid state physics to describe the constituent. Although the field concerned is now a scalar (i.e. Pauli exclusion principle does not apply), the $\psi$ field has a constant density on symmetry breaking as a condensate and therefore any defects on the field will behave as “holes” in such condensate. If the potential term is a constant $V_0$ at the ground state of the space-time phase when it is filled up with the $\psi$ field, then the lagrangian of a defect or “holes” of the $\psi$ field on the manifold can be written as

$$L_h = \partial_\mu \psi_h \partial^\mu \psi_h + \frac{R_0}{12}\psi_h^2 - \frac{\lambda}{36}\psi_h^4 = V_0 - (\partial_\mu \psi \partial^\mu \psi + \frac{R_0}{12}\psi^2 - \frac{\lambda}{36}\psi^4) \quad (5)$$

where $\psi_h$ is the wavefunction for the holes and therefore we can write the equation as

$$S_g = \int (\frac{1}{2}\partial_\mu \psi_h \partial^\mu \psi_h + \frac{1}{12}R_0\psi_h^2 - \frac{1}{36}\lambda\kappa\psi_h^4 - V_0)\sqrt{-g}d^4x \qquad (6)$$

where

$$\mu^2 / 2 = R_0 / 12 \text{ and } g/4 = \lambda\kappa / 72 \qquad (7)$$

The lagrangian in equation (6) can be identified as the usual scalar field in the electroweak theory where $\mu$ and $g$ are related as $\mu^2 / g = v^2$ and also, $V_0$ can be assumed to be zero. We have to remark that the above identification is true only when $R$ is independent with $g_{\mu\nu}$ and be a constant $R_0$ as assumed in extreme microscopic scale. Therefore, the $\lambda$ appears here is not representing the macroscopic observed cosmological constant. Equation (7) gives the relations that bring together GR with the electroweak theory. We can further find that

$$R_0 = \frac{2\nu^2 \lambda \kappa}{3} \tag{8}$$

From the above calculation, we know that GR and electroweak theory could be the same theory on different length scale and $R_0$ is directly related to $\mu^2$. $R_0$ can be interpreted as the curvature of the fundamental space-time unit. If we put the symmetry broken field $\psi_h = \nu$ into the potential term of equation (6), we get

$$\langle V \rangle = \rho = -\frac{g\nu^4}{4} \tag{9}$$

This value represents the vacuum energy density of the space-time fundamental unit when the symmetry is broken at electroweak energy scale. It is not the value in cosmological observation since we have to weighed it through the continuous space as described in the previous paper [1]. From the discussion of equation (5), we know that the "holes" of the $\psi$ field represents the space-time constituent and is denoted as $\psi_h$. If the probability density of the $\psi$ field can be written as $\rho_\psi = 2q\psi^2$ as usual in scalar field (where $q$ is the energy of a scalar particle), then the probability for the "holes" field $\psi_h$ is as $\rho_{\psi_h} = 2q{\psi_h}^2$. Since the $\psi_h$ field represents the wavefunction of the space-time constituent, the number density of the space-time contituent is given by the $\rho_{\psi_h}$ expression. As in the Dirac's holes theory for positron, we assume that the "holes" for the constituent is due to a defect of the state of the $\psi$ field with negative energy state. So that we let $q < 0$ and therefore $\rho_\psi > 0$. We can further write the energy density constituent of the $\rho_{\psi_h E}$ as $\rho_{\psi_h E} = -2|q|^2 {\psi_h}^2$. The vacuum $\psi_h$ field is a constant on spontaneous broken symmetry and this makes the number density of the space-time constituent a constant too. By using the energy density calculated above in equation (9) $\rho_{\psi_h E} = -g v^4 / 4$ and putting $\psi_h = v$, we find that $|q| = v\sqrt{g/8}$ and the number density is then as $\rho_{\psi_h} = v^3\sqrt{g/2}$. The energy $q$ behaves as the binding energy of the constituent but it is not equal Higgs mass because energy gap may exist between the vacuum and the Higgs excitation state of the constituent. It is interesting to note that if the Higgs mass is about $115\,GeV$ (although not yet confirmed in the LEP experiment of CERN), the energy gap can be found as $86GeV$ ( $g = 0.109$ is used) which is about the average mass of the W & Z particle. By the most fundamental hypothesis of our theory, the dimension of the space-time constituent is of the order of Planck scale so that the weighing factor for converting the microscopic vacuum energy density to macroscopic case can then be found as $(v / kM_p)^3\sqrt{g/2}$ where $k$ is a geometrical factor. Therefore, the macroscopic cosmological constant $\lambda_m$ can be written as

$$\lambda_m = \frac{\kappa g v^7}{4(kM_p)^3}\sqrt{g/2} = \frac{\kappa\sqrt{g^3}\, v^7}{4\sqrt{2}(kM_p)^3} \sim \frac{m^7}{M_p^5} \qquad (10)$$

where $m$ is the electroweak mass scale and can be taken as about $100GeV$. This verifies the estimation in the previous paper [1]. We have to introduce the factor $k$ here because the actual size of the space-time constituent is still unknown and it can represent the radius of the constituent in faction of Planck scale. However, we will then show that such information is already in GR.

By Newtonian approximation of GR, we know that $R \sim -\partial^2\varphi = -\partial_\mu E^\mu = 4\pi\rho / M_p^2$ where $R$ is the Ricci scalar, $\varphi$ is the gravitational potential and $E$ being the graviational field. Using Stoke's theorem, we can write $\partial_\mu E^\mu = \oint_S E \cdot dS / dV$. If we consider a electroweak scale spherical volume element of radius $r$ that cover a space-time constituent and the field $E$ is spherical symmetrical on it, we can find that $\partial_\mu E^\mu = 3E / r$ ($E$ is the field experience by the constituent from the nearby constituent.). If we use the previous result that the density of the space-time constituent is $\rho = \nu^3 \sqrt{g/2}$ and rectangular volume is assumed, we find that the spacing of the constituent is $r = 1/1.23\nu$. Therefore, we get

$$R \sim \bar{o}_\mu E^\mu = 3 \times 1.23 E\nu = 3.69 E\nu \qquad (11)$$

Also, from the above results, the equivalent energy of a space-time constituent is $q = \nu\sqrt{g/8}$ and since the stress force experience by the constituent $F$ is equal to $Eq$, the above equation (11) can be expressed as

$$R \sim 3.69 F\sqrt{8/g} \qquad (12)$$

The vacuum energy $\rho_v$, vacuum pressure $P$ and surface area of the constituent as $4\pi r_p^2$ are related to $F$ as $F = -4\pi r_p^2 P = 4\pi r_p^2 \rho_v$ where $r_p$ is the radius of the constituent. So that equation (12) becomes

$$R \sim 3.69 \times 4\pi r_p^2 \rho_v \sqrt{8/g} = \frac{4\pi\rho_v}{M_p^2} \qquad (13)$$

The value $r_p$ can be solved as $r_p = 1/(5.62 M_p)$ and therefore the $k$ value can be found as $5.62$. This also shows that the gravitational constant is directly related to the size of the space-time constituent. If we put this value back into equation (10), the expression for the cosmological constant can then be found as

$$\lambda_m \sim 10^{-4} \frac{\pi\nu^7}{M_p^5} = 0.15 \frac{m^7}{M_p^5} \sim 10^{-83} GeV^2 (14)$$

The value given in equation (14) is close to the observation value by one order of magnitude. This one order discrepancy could be due to the uncertainty in the geometrical assumption, the deviation from Euclidean geometry in extreme small scale or some other unknown reason. Therefore, further investigation should be needed on this point. On the other hand, we can treat the problem in a reverse way. If we assume that the size of the constituent is unknown, we can then determine it by the cosmological observation of the $\lambda_m$ value. It provides the ground for an experimental determination of the size of the space-time constituent by the cosmological observation of the $\lambda_m$. This is a very impressive point in our theory.

Besides finding the value of the cosmological constant, the above calculation also explains the weakness of the gravitational constant. We can observe that the gravitational constant acts like the interaction cross section of vacuum energy on the space-time constituents. This makes its dimension as a cross sectional area (using $\hbar = c = 1$). Its weakness is due to the extremely small cross sectional area of the constituent when the space-time is assumed to be discrete. Since its origin is the size of the constituent, its nature may be more fundamental than other coupling constant. We can now say that the dimension of the gravitational constant may not the major barrier to the quantization of gravity or make it renormalizable. As we mentioned in the previous paper, the discrete nature of space-time makes the metric becomes a kind of collective field and the coupling constant becomes a cross sectional area. This situation makes the direct quantization of the metric meaningless. In this point of view, previous method of quantum gravity may be in the wrong direction and not straight to the problem. We propose that a quantum theory of gravity should be based on the quantization of the space-time constituent field since it is the actual fundamental field of the space-time.

We now have a more theoretical treatment on the reason of why the electroweak scale and the Planck scale can enter the expression of the cosmological constant by the discrete space-time hypothesis. Although most of the physical meaning has been introduced in the previous paper [1] but now we have a more formal and mathematical treatment on it. In the mathematical process, we know that the lagrangian for Higgs scalar and GR can actually be combined into a single theory and the field for the space-time constituent is just the scalar field in electroweak theory. This explains the physical meaning of the existence of a scalar field in the vacuum and why such field can take the role in the inflation theory [10]. In more detail physical description, the space-time constituents can be interpreted as the “holes” embedded in the “sea” of the $\psi$ field which exerts negative pressure on such “holes” and gives the cosmological constant. If the energy density increase to above the electroweak energy scale and the space-time phase transition happens, large amount of constituents will be excited as Higgs particle and left with high density of “holes”. This makes the corresponding cosmological constant has a large increase as described in our previous paper. This process acts like external energy is absorbed to become the vacuum energy. In the reverse cooling process, Higgs particle loss its energy and combined with the “holes” by creating other particle with positive energy and the cosmological constant decreases due to lower density of holes. This implies the relation between the matter energy density and the vacuum energy density during the phase transition as $\Delta\rho_{vac} + \Delta\rho_{matter} = 0$ and verifies our description in our previous paper. The weakness of the gravitational constant can also be explained by the smallness size of the constituents as a consequence. It enters the gravitational field equation as the interaction cross section of the matter density to the constituents. In view of the theory, the phenomenon of accelerating universe may hint us that space-time is discrete in nature and the information of the space-time microscopic structure can be obtained from cosmological observations. This point is impressive but not too unfamiliar to us since previous scientists had also developed the atomic theory of matter from macroscopic phenomenon, such as kinetic properties of gas and diffusion. Also, from the experience of atomic theory and quantum theory of black body, theoretical infinities may imply some discrete nature. We propose that it may also be true for space-time in quantum gravity. At least, it is a viable direction for us to pursue further.

**Reference**

[1] H.M.Mok, astro-ph/0105513

[2] M.S.Turner, astro-ph/0108103

[3] S.M.Carroll, astro-ph/0107571

[4] S.Weinberg, astro-ph/0104482

[5] G.t'Hooft, Recent Development in Gravitation, edited by M.Levy & S.Denser, (Plenum, New York, 1979); gr-qc/9608037; gr-qc/9601014

[6] S.Perlmutter *et al*., Astrophys. J. **517**, 565 (1999)

[7] B.P.Schmidt *et al*., Astrophys. J **507,** 46 (1998)

[8] A.G.Riess *et al*., astro-ph/0104455

[9] S.Weinberg, Rev. Mod. Phys. **61** 1 (1989)

[10] A.Linde, Particle Physics and Inflationary Cosmology (Harwood Academic, GmbH, 1990)